\documentclass[aps,pre,twocolumn,showpacs,superscriptaddress]{revtex4-2}
\usepackage{graphicx}  
\usepackage{dcolumn}   
\usepackage{bm}        
\usepackage{dsfont}    
\usepackage{amssymb}   
\usepackage{amsmath}
\usepackage{journals}
\usepackage{color}
\usepackage{siunitx}
\usepackage{physics}
\usepackage{hyperref}

\usepackage[utf8]{inputenc}

\begin{document}
\title{Exposure theory for learning complex networks with random walks}
\author{Andrei A.~Klishin}
\affiliation{Department of Bioengineering, University of Pennsylvania, 
Philadelphia, PA 19104 USA}
\author{Dani S.~Bassett}
\email{dsb@seas.upenn.edu}
\affiliation{Department of Bioengineering, University of Pennsylvania, 
Philadelphia, PA 19104 USA}
\affiliation{Department of Physics \& Astronomy, University of Pennsylvania, 
Philadelphia, PA 19104 USA}
\affiliation{Department of Electrical \& Systems Engineering, University of 
Pennsylvania, Philadelphia, PA 
19104 USA}
\affiliation{Department of Neurology, University 
of Pennsylvania, Philadelphia, PA 19104 USA}
\affiliation{Department of Psychiatry, University 
of Pennsylvania, Philadelphia, PA 19104 USA}
\affiliation{Santa Fe Institute, Santa Fe, NM 87501 USA}

\date{\today}

\begin{abstract}
	Random walks are a common model for exploration and discovery of complex 
	networks. While numerous algorithms have been proposed to map out an 
	unknown network, a complementary question arises: in a known network, which nodes and edges are most likely to be discovered by a random walker in finite time? Here we introduce exposure theory, a statistical mechanics framework that predicts the learning of nodes and edges across several types of networks, including weighted and temporal, and show that edge learning follows a universal trajectory. While the learning of individual nodes and edges is noisy, exposure theory produces a highly accurate prediction of aggregate exploration statistics.
\end{abstract}

\maketitle

\section{Introduction}
Random walks are a common baseline model of dynamical processes on 
networks, representing phenomena from search to disease spreading to 
communication \cite{masuda2017random}. The problem of random walk statistics, such as the number of visited nodes and the return probability, was first formulated and solved on infinite, periodic, low-dimensional lattices \cite{dvoretzky1951some, polya1921aufgabe}, but the early numerical studies on finite, complex, irregular networks revealed quite different exploration behaviors \cite{costa2006learning, costa2007exploring, asztalos2010network}. Along with node- and edge-level exploration statistics, many formal results have been obtained for aggregate metrics, such as cover and return times \cite{volchenkov2011random,cooper2007sparse,cooper2007pref,maier2017cover}. Random walk statistics can be used to infer the underlying network structure, from node centrality \cite{delvenne2011centrality} to community structure \cite{rosvall2008maps,piccardi2011finding}. While simple random walks independently revisit known nodes and edges, a variety of modified random walk algorithms have been developed to speed up network exploration or deliberately revisit the known parts \cite{costa2006learning, asztalos2010network, delvenne2011centrality, sinatra2011maximal, dearruda2017knowledge, iacopini2018network}.

Beyond their relevance for dynamical network processes in the environment, random walks are also relevant for cognitive processes taking place in the human mind. Recent experiments show that humans (and machines) can learn 
network structures from observing random walks taken upon them \cite{schapiro2013neural, stachenfeld2017hippocampus, lynn2020review, karuza2021value}. Unlike machines, humans commit mental errors in learning networks, which serves to foreground large scale network structure over fine grained details \cite{lynn2020errors}. The dual needs to minimize the impact of such errors while maximizing the communication efficiency might serve as selection pressures on communication network architectures \cite{lynn2020human}.

Our goal is to understand how the learning of complex networks by random walks is driven by the interplay of the finite length of random walks and the structure of the learned networks. In order to quantify this interplay, in this paper we propose and validate \emph{exposure theory}, a framework similar to the conventional Boltzmann statistical mechanics in that it predicts a local probability of visiting an individual edge and provides a rule for combining such probabilities to predict aggregate statistics. We validate both local and aggregate predictions against direct stochastic random walk simulations and find exposure theory to be highly accurate at a fraction of computational cost. Exposure theory systematizes and generalizes a variety of previous results on random walks, and can be further expanded to account for human information processing.

\section{Mathematical formalism}
\begin{figure}
	\includegraphics[width=\columnwidth]{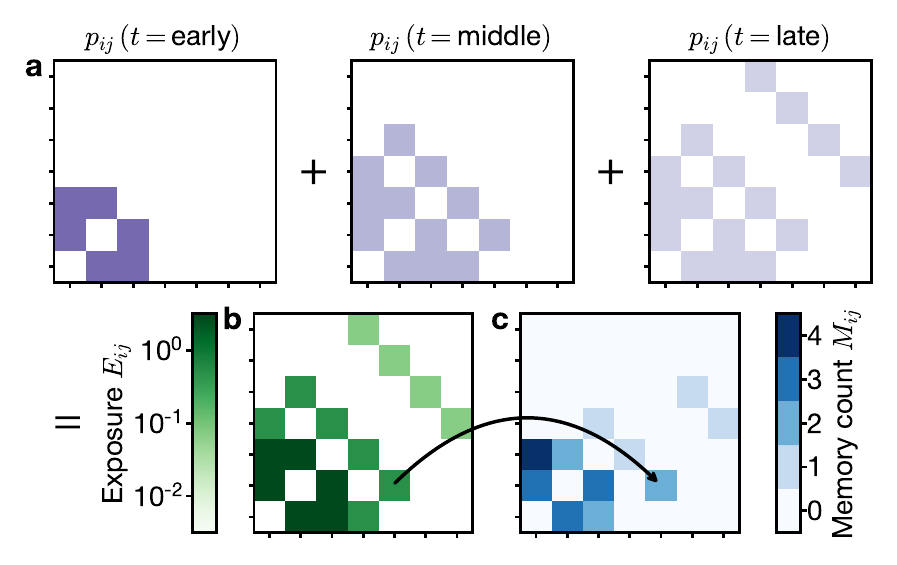}
	\caption{\textbf{Exposure theory consists of a deterministic accumulation of exposure and a stochastic draw of memory count.} (a) As the network grows, at every time point $t$ we 
	compute the matrix of edge visit probabilities $\mathbf{p}(t)$ and sum them up to obtain the integral exposure matrix $\mathbf{E}$ (b). (c) Each entry of the exposure matrix is the parameter of an independent 
	Poisson distribution from which we draw the integer entries of the memory 
	count matrix $\mathbf{M}$.}
	\label{fig:cartoon}
\end{figure}

We consider random walks on an undirected, weighted, time-dependent network described by the adjacency matrix $\mathbf{A}(t)$. The \emph{learning} of the network is described by the integer-valued memory matrix $\mathbf{M}$: every time the walk traverses (visits) the edge $(i,j)$, we add 1 to the 
element $M_{ij}$. How are the different elements of $\mathbf{M}$ populated over a finite time? On one side, we simulate several random walk realizations and compute the statistics. On the other side, we introduce the exposure ensemble that predicts 
the memory distribution directly, but approximately.

We first compute the steady-state probability of visiting a particular edge $p_{ij}$. In random walks on a weighted network the conditional probability $P(j|i,t)$ of the step $i\to j$ is given by the row-normalized adjacency matrix. We combine it with the steady-state probability of ending up on a particular node $\pi_i(t)$ so that the row norm cancels out:
\begin{align}
	P(j|i,t)=\frac{A_{ij}(t)}{\sum_j A_{ij}(t)};&\quad 
	\pi_i(t)=\frac{\sum_j A_{ij}(t)}{\sum_{ij} A_{ij}(t)};\\
	p_{ij}(t)=\pi_i(t) P(j|i,t)&=\frac{A_{ij}(t)}{\sum_{ij}A_{ij}(t)}.
\end{align}

What is the probability of visiting the edge $(i,j)$ multiple times and thus getting multiple memory counts $M_{ij}$? Technically, subsequent visitations of the same edge are not conditionally independent. The conditional dependence is especially strong in lattices \cite{dvoretzky1951some}, in Watts-Strogatz networks derived from 1D lattices \cite{almaas2003scaling}, and other networks embedded in a low-dimensional space \cite{maier2017cover} (see Supplementary Materials (SM) for explicit test). However, many other complex networks do not have a low-dimensional latent space and thus have very short mixing or relaxation times \cite{maier2017cover}. For such networks, we can build exposure theory resting on three key assumptions: subsequent edge visitations are conditionally independent, their probability distribution follows the instantaneous steady-state $p_{ij}(t)$, and $p_{ij}\ll 1$. In this case the accumulation of memories of each edge follows an independent, possibly non-stationary Poisson process (see Fig.~\ref{fig:cartoon} and SM):
\begin{align}
	M_{ij}(t)\sim \textsf{Pois}(E_{ij}(t));\quad E_{ij}(t)\equiv 
	\sum\limits_{1}^{t} p_{ij}(t'),
	\label{eqn:exposure}
\end{align}
where we defined the quantity $E_{ij}$ as the \emph{integral exposure} of the edge $(i,j)$. The name is inspired by the operation of a camera: the shutter is opened for 
a specific time $t$, during which the film or sensor inside integrates the incoming flux of light. The entries $E_{ij}$ form a matrix that depends on the length of the random walk.

How does the exposure matrix relate to properties of a network, such as its weight distribution and time-dependence? We first consider a time-independent network, in which the visitation probability is also time-independent. In this case, the sum in Eq. \eqref{eqn:exposure} consists of $t$ identical terms and is trivially computed to yield:
\begin{align}
	E_{ij}(t)=t\cdot \mathcal{E}_{ij};\quad \mathcal{E}_{ij}\equiv p_{ij}= 
	\frac{A_{ij}}{\sum_{ij}A_{ij}},
\end{align}
where $\mathcal{E}_{ij}$ is the \emph{specific exposure} (per random walk step). Thus for time-independent networks the exposure of each edge grows linearly in time in proportion to the relative edge weight. For unweighted networks, the exposure of all edges is the same $E_{ij}=t/m$, where $m$ is the number of edges.

In contrast, if the underlying network is time-dependent, then we assume that it evolves under some deterministic protocol $A_{ij}(\tau)$ at the same time as the random walk unfolds. We synchronize the random walk time $t$ with the evolution time $\tau$ by setting their ratio to
$t/\tau\equiv D$, and refer to this quantity as the \emph{dilation}. A low value of the dilation parameter indicates that network evolution is fast compared to the random walk, whereas a high value indicates that the random walk can explore the network before it changes significantly. In this case the exposure can be computed as follows:
\begin{align}
	E_{ij}(t)=D\cdot \mathcal{E}_{ij}(t/D);\quad \mathcal{E}_{ij}(\tau)\equiv 
	\int\limits_{0}^{\tau} p_{ij}(\tau')d\tau',
\end{align}
where $\mathcal{E}_{ij}(\tau)$ is the time-dependent version of specific exposure. Varying the value of dilation $D$ thus rescales the absolute value of integral exposure, but stretches out its evolution.

The specific exposure matrix $\mathbf{\mathcal{E}}$ is another way to represent a network with a matrix, alongside others such as adjacency, Laplacian and 
modularity matrices \cite{newman2006modularity}. While other matrices do not change their meaning significantly when all elements are uniformly rescaled by a scalar, such rescaling is crucial in converting from specific to integral exposure. The integral exposure matrix describes the \emph{canonical ensemble} of
random walk memories over the network since the total number of memory counts
fluctuates around the average of $t$ (see SM). In a conventional Boltzmann canonical ensemble the probability of a state $p(s)\propto e^{-\beta H(s)}$ is driven by the product of the inverse temperature $\beta$ and the Hamiltonian $H(s)$; in the exposure ensemble the memory distribution (Eq.~\eqref{eqn:exposure}) is driven by the product of the prefactor $t$ or $D$ and the edge specific exposure $\mathcal{E}_{ij}$. While the specific probability formula differs, a range of computational techniques carry over, especially for coarse-graining (see SM). 

Similarly to the Boltzmann ensemble, we are interested in the limiting cases of large and small exposure. The large exposure limit is easily reached by making the prefactor $t$ or $D$ large. In this case the Poisson distribution is well-approximated by its mean, and thus the number of counts of each edge is nearly equal to its exposure. The small exposure limit is more subtle. The probability of having more than zero memories of a particular edge is given by the expression:
\begin{align}
	P(M_{ij}>0)=1-e^{-E_{ij}(t)},
\label{eqn:visit}
\end{align}
which implies that the probability of learning any edge \emph{locally} follows a universal exponential curve in terms of its exposure. 



For small learning times, the learning of individual edges remains highly noisy. However, since the learning of each edge is independent, we expect to get much more 
accurate results by aggregating across multiple edges. In particular, exposure theory can predict the \emph{fraction of edges} that have been learned:
\begin{align}
	V(t)=1-\frac{1}{m}\sum\limits_{ij\in A_{ij}}e^{-E_{ij}(t)}\leq 
	1-e^{-t/m},
	\label{eqn:fraction}
\end{align}
where the sum runs over the edges that exist in the adjacency matrix. The bound comes from applying Jensen's inequality to the sum (see the SM). The right hand side 
expression corresponds to the fraction of edges learned in an unweighted 
network of $m$ directed edges. Therefore, Jensen's inequality imposes a 
fundamental limit on the speed of learning, regardless of the distribution of 
network weights and the time-dependent protocol for network change. The predictions of both the detailed shape of $V(t)$ and the Jensen bound can also be tested by direct comparison to stochastic random walk simulations.

\section{Validation}

\begin{figure}
	\includegraphics[width=\columnwidth]{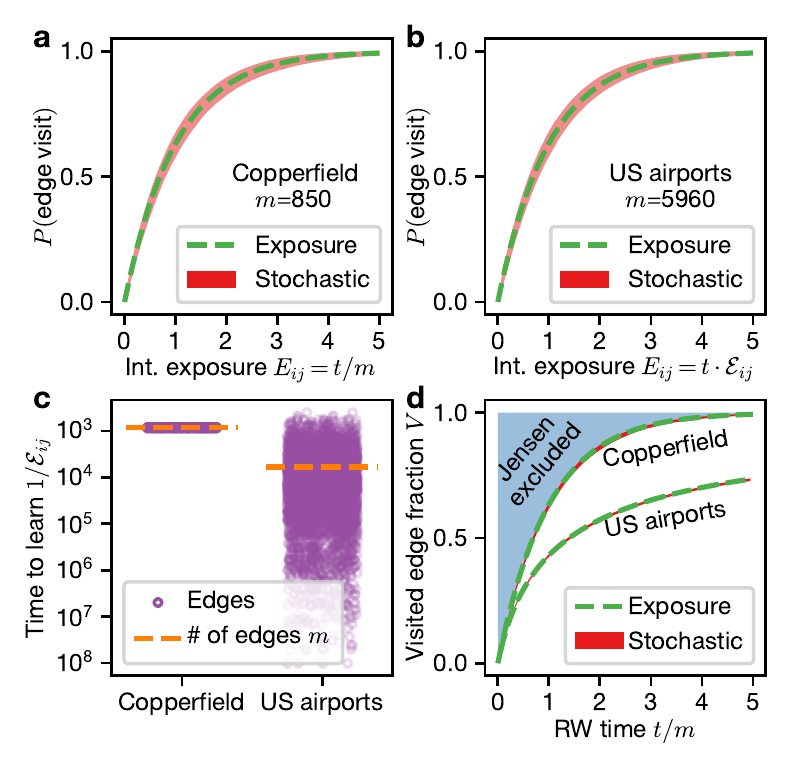}
	\caption{\textbf{Validation of local and aggregate exposure theory predictions for time-independent networks.} (a-b) Universal curve of edge visit probability (green dashed) 
	describes the stochastic probability over 100 runs (red shaded, 
	mean$\pm$std) averaged over $m$ directed edges. Time is scaled by edge 
	exposure: uniform for the unweighted network (a) and edge-specific for 
	the weighted network (b). (c) Scatterplot of time to learn each edge in the unweighted and 
	weighted networks. (d) Visited edge fraction against time scaled by the number of edges. The blue region is excluded by the Jensen bound, which the unweighted network saturates.}
	\label{fig:probs}
\end{figure}

Exposure theory has yielded two predictions: a local learning curve (Eq.~\ref{eqn:visit}) and an aggregate exploration curve (Eq.~\ref{eqn:fraction}). While these predictions assume conditional independence of visitations, we validate them with stochastic simulations that faithfully capture the conditional dependence on two static networks and one dynamic network. The 
first network (Copperfield) tracks the co-occurrence of nouns and adjectives in 
the text of \emph{David Copperfield}, a novel by Charles Dickens, and is 
unweighted and undirected \cite{newman2006finding}. The second network (US 
airports) tracks the passenger flow between the top-500 busiest airports 
in the United States \cite{colizza2007reaction}. The passenger flow varies by a 
factor of $10^5$ across the edges, but is symmetric within each pair, resulting 
in a weighted undirected network.

\begin{figure*}
	\includegraphics[width=\textwidth]{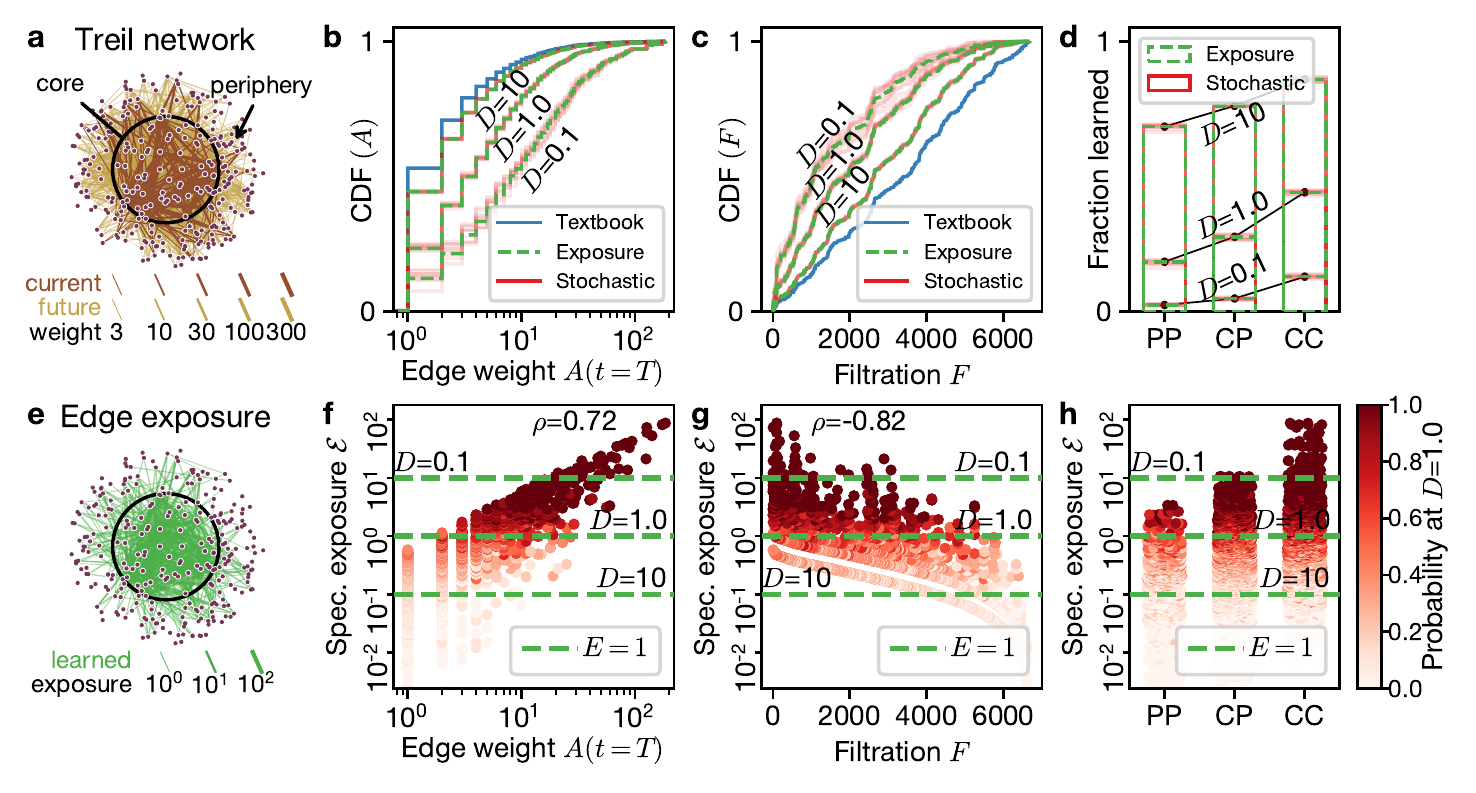}
	\caption{\textbf{Exposure theory predicts the architecture of learned networks.} (a) The Treil network edges vary widely in both edge weight and 
	filtration order, and together form a robust core-periphery structure. (b-c)
	Cumulative distribution functions (CDFs) for edge weight by the end of the book ($A(t=T)$, b) and 
	filtration order ($F$, c) for the textbook network (blue), stochastic 
	simulations (red), and exposure prediction (green). (d) Fraction of edges 
	learned in different structural positions: within-core (CC), core-periphery 
	(CP), and within-periphery (PP). (e) The Treil network with edges valued by exposure. (f-h) Scatter plots of all edges in the	network, with the color corresponding to the probability of the edge being
	learned across 10 replicas at $D=1.0$. The green horizontal dashed lines 
	show the condition $E=D\mathcal{E}=1$, which is a good separator of 
	learned and non-learned edges. Panels (f,g) include the 
	Spearman correlation coefficient $\rho$ between specific exposure $\mathcal{E}$ and edge weight $A(t=T)$ and filtration $F$, respectively, $\log_{10}(p)<-12$.}
	\label{fig:exposure}
\end{figure*}

On each static network, we simulate 100 replicas of stochastic random walk realizations that begin at a randomly chosen node. In order to test the local prediction, for each edge we compute the fraction of replicas in which it was visited by time $t$. We convert time $t$ into integral exposure by rescaling it: for the Copperfield network, the rescaling is uniform for each edge $t\to t/m$, whereas for the US airports network 
the rescaling is edge-specific $t\to t\cdot \mathcal{E}_{ij}$. After rescaling, 
we average the probability across the edges in each network. In 
both cases (Fig.~\ref{fig:probs}a,b) the prediction of Eq.~\ref{eqn:visit} 
lines up with the stochastic results extremely well, which points to the universal nature of edge learning and several practical corollaries. First, learning edges in the pair $(i,j)-(j,i)$ is equally likely but independent. Second, in unweighted networks such as Copperfield, learning all edges is equally likely. Third, the characteristic time to learn any edge is $1/\mathcal{E}_{ij}$, which varies over many orders of magnitude (Fig.~\ref{fig:probs}c).

After the local predictions, we test the aggregate prediction of the visited edge fraction $V(t)$. The learning of the unweighted Copperfield network exactly saturates the Jensen bound, while the weighted US airports network significantly underperforms (Fig.~\ref{fig:probs}d). In both cases the variance of $V$ is extremely small, which points to the accuracy of exposure theory predictions of aggregate metrics. Similar functional forms of the aggregate learning curve have been previously observed or assumed without a derivation 
\cite{costa2006learning, asztalos2010network, dearruda2017knowledge}.

Having confirmed the universal learning curve on static networks, we extend exposure theory predictions to a temporal network that we adapted from 
Ref.~\cite{christianson2020architecture}. The authors represented several 
popular linear algebra textbooks as temporal networks, where nodes are key 
concepts and edges are co-occurrence of concepts in a sentence. The 
time-dependent adjacency matrix is defined as 
$A_{ij}(\tau)=A_{ij}\cdot[F_{ij}<\tau]$, where $A_{ij}$ is the number of co-occurrences of the pair $(i,j)$, $F_{ij}$ is the filtration order (time of first co-occurrence), $[\cdot]$ is an indicator function, and time 
$\tau\in[0,\tau_{max}]$ is measured in sentences of the book. Out of the 10 textbooks, here we analyze the network of the textbook by Treil 
(Fig.~\ref{eqn:exposure}a).

We compare the taught and the learned networks by three architectural 
metrics: edge weight $A$, filtration order $F$, and core-periphery structure. The cumulative distribution functions (CDFs) for the learned edges are biased towards higher $A$ and lower $F$ with respect to the textbook, but the bias decreases with dilation $D$ (Fig.~\ref{eqn:exposure}b,c, see SM for details). Edges within the core (CC) are most likely to be learned, followed by core-periphery (CP) and within-periphery (PP) edges (Fig.~\ref{eqn:exposure}d). In all cases exposure theory provides an excellent match to the stochastic CDFs.

Finer architectural details of learned networks are revealed through the lens of edge-level specific exposure (Fig.~\ref{eqn:exposure}e). The specific exposure of individual edges $\mathcal{E}_{ij}$ depends locally on the edge weight $A_{ij}$, but also non-locally on the sum of all edge weights $\sum_{ij}A_{ij}(t)$ and temporally on the filtration order $F_{ij}$ (Spearman correlation in Fig.~\ref{eqn:exposure}f-g). While edge weight $A_{ij}$ for the Treil network spans two orders of magnitude, the range of specific exposure stretches out to nearly six orders of magnitude. As dilation $D$ increases, more edges cross the exposure threshold $E_{ij}=D\mathcal{E}_{ij}\simeq 1$ and are learned (Fig.~\ref{eqn:exposure}f-h). Specific exposure predicts not only the binary learning outcome, but also detailed statistics of the memory count $M_{ij}$ at any dilation (see SM).

\section{Discussion}
The proposed exposure theory builds upon several previous results, but simultaneously generalizes them under a single statistical mechanics framework. Exposure theory improves upon previous results in three aspects. First, we center our analysis on the network edges and obtain results for nodes as a simple corollary via a coarse-graining process (see SM). Second, we focus on the probability distributions and time dependence of quantities of interest as opposed to just the means. Third, we consider not just one but all visitations of an edge or a node and thus do not need to introduce artificial absorbing states \cite{masuda2017random}. Exposure theory relies on seemingly stringent assumptions of network properties, but these assumptions hold for most real-world complex networks of non-spatial origin, as discussed in Refs.~\cite{maier2017cover, lau2010asymptotic} and as shown in the context of exposure predictions in the SM.

The two key predictions of exposure theory, the local visit probability (Eq.~\ref{eqn:visit}) and the aggregate visit fraction (Eq.~\ref{eqn:fraction}), look deceptively similar mathematically but have been analyzed separately before. The time to visit an edge or node is also commonly known as the first passage time, and the visitation probability in Eq.~\ref{eqn:visit} serves as its cumulative distribution function. Usually interest in the \emph{mean} first passage time stimulates the development of computational methods \cite{bartolucci2021spectrally}, but Ref.~\cite{lau2010asymptotic} argued that the first passage time \emph{distribution} has a universal exponential tail with variable decay rate consistent with our Eq.~\ref{eqn:visit}. The decay rates of the first passage time distribution can be used to compute the cover time, or average time to learn the whole network \cite{maier2017cover}. The aggregate visit fraction in Eq.~\ref{eqn:fraction} tracks not only the time to visit all edges or nodes, but also the detailed dynamics of approaching that limit, thus providing a theoretical backing to the exploration curves reported in Refs.~\cite{costa2006learning, asztalos2010network}. In addition to deriving the exploration curve, we provide a fundamental Jensen bound on the speed of exploration possible with random walks.

Random walks are a paradigmatic example of a dynamical process used to map out \emph{a priori} unknown networks. However, the properties of such a dynamical process can produce a significantly distorted view of network architecture. For example, when the physical topology of the Internet is studied with the common traceroute algorithm, the measured network can have a spurious power-law node degree distribution even if the underlying network has narrowly distributed or even identical node degrees \cite{clauset2003traceroute, achlioptas2009bias}. Acknowledging this distortion led to the development of unbiased sampling techniques \cite{stutzbach2008unbiased}. Exposure theory accounts for the distortions produced by the under-sampling bias of finite random walks (Fig.~\ref{fig:exposure}). While at the binary level (existence of nodes or edges) the sampling bias decreases and disappears with longer walk time, it remains to be seen if more subtle distortions of edge weight or node strength persist for longer times.

Apart from learning the detailed network topology, finite-time spreading processes such as random walks have been used to infer a variety of other structural metrics. The spreading time can serve as a natural multiscale lens to study network structure from community partitions \cite{lambiotte2008laplacian, schaub2012markov} to node centrality \cite{arnaudon2020scale}. Finite-time spreading is also key to the quantum-like entanglement of network flows \cite{ghavasieh2020statistical, ghavasieh2021statistical}. Random walks inspire centrality measures for multiplex networks \cite{rahmede2018centralities}.
Hyperbolic network models allow tuning the network spectral dimension \cite{bianconi2017emergent}, which in turn drives the behavior of dynamical processes such as synchronization and diffusion \cite{millan2019synchronization, millan2021local}. Lastly, random walks are key to the diffusion part of reaction-diffusion systems \cite{hufnagel2004forecast, colizza2007reaction}, which in turn are crucial to temporal prediction of epidemic spreading and related processes \cite{brockmann2013hidden, iannelli2017effective}.

\section{Conclusions}
In this paper we studied random walk exploration of a broad class of complex networks that span from unweighted, undirected to weighted and temporal. For these networks, our proposed exposure theory produces detailed and accurate predictions from edge-level probability to aggregate learning curves, with no free parameters beyond those required to specify the random walk conditions. Exposure computations are orders of magnitude faster than stochastic simulations (see SM for benchmarks). While the speed of exploration by simple random walks is limited by the Jensen bound, it can be surpassed by biased random walks \cite{bonaventura2014characteristic}, intelligent exploration algorithms \cite{asztalos2010network}, or multiple interacting walkers \cite{dearruda2017knowledge}. Exploration dynamics might also deviate from the universal curve for directed networks that support more complex steady states \cite{costa2006learning}, or temporal networks that connect and disconnect components dynamically \cite{holme2012temporal, perra2012random, petit2019classes}. Even for simple random walks, the learned networks can be significantly distorted by the faults of human memory \cite{lynn2020errors}; we are looking forward to extensions of exposure theory to explore such distortions in detail.

\section*{Acknowledgments}
The authors would like to thank N.~Christianson for providing the linear algebra textbook network data, and C.W.~Lynn and X.~Xia for discussions about the modeling. The computational workflow in general and data management in particular for this work was primarily supported by the Signac data management framework \cite{signac_commat, signac_zenodo}. This research was funded by the Army Research Office award number DCIST-W911NF-17-2-0181 and the National Institutes of Mental Health award number 1-R21-MH-124121-01.

\section*{Citation Diversity Statement}

Recent work in several fields of science has identified a bias in citation practices such that papers from women and other minority scholars are under-cited relative to the number of such papers in the field \cite{mitchell2013gendered, dion2018gendered, caplar2017quantitative, maliniak2013gender, dworkin2020extent, bertolero2021racial, wang2021gendered, chatterjee2021gender, fulvio2021imbalance, teich2021citation}. Here we sought to proactively consider choosing references that reflect the diversity of the field in thought, form of contribution, gender, race, ethnicity, and other factors. First, we obtained the predicted gender of the first and last author of each reference by using databases that store the probability of a first name being carried by a woman \cite{dworkin2020extent, zhou_dale_2020_3672110}. By this measure (and excluding self-citations to the first and last authors of our current paper and references in this paragraph), our references contain 6.25\% woman(first)/woman(last), 8.33\% man/woman, 14.58\% woman/man, and 70.83\% man/man. This method is limited in that a) names, pronouns, and social media profiles used to construct the databases may not, in every case, be indicative of gender identity and b) it cannot account for intersex, non-binary, or transgender people. Second, we obtained predicted racial/ethnic category of the first and last author of each reference by databases that store the probability of a first and last name being carried by an author of color \cite{ambekar2009name, sood2018predicting}. By this measure (and excluding self-citations and references in this paragraph), our references contain 10.9\% author of color (first)/author of color(last), 20.54\% white author/author of color, 16.8\% author of color/white author, and 51.76\% white author/white author. This method is limited in that a) names and Florida Voter Data to make the predictions may not be indicative of racial/ethnic identity, and b) it cannot account for Indigenous and mixed-race authors, or those who may face differential biases due to the ambiguous racialization or ethnicization of their names.  We look forward to future work that could help us to better understand how to support equitable practices in science.

\bibliography{biblio,bibfile_CDS}

\end{document}


\title{Exposure theory for learning complex networks with random walks:\\
Supplementary Materials}
\author{Andrei A.~Klishin}
\author{Dani S.~Bassett}
\email{dsb@seas.upenn.edu}

\date{\today}


\maketitle
\section{Derivation of exposure theory}
Here we provide a first-principles derivation of exposure theory from the three assumptions stated in the main text:
\begin{enumerate}
    \item Subsequent edge visitations are conditionally independent from each other.
    \item Probability distribution of visitation follows the instantaneous steady-state $p_{ij}(t)$.
    \item Probability of visiting a particular edge in one step is small $p_{ij}(t)\ll 1$.
\end{enumerate}

\noindent These assumptions are quite frequently satisfied for complex networks studied in this paper. Assumptions 1 and 2 rely on fast mixing (short correlation time) of random walks, which we explicitly compute for the studied networks in Section~\ref{sec:correlation}. Assumption 1 additionally implies that the network is connected---otherwise the probability of visiting an edge in one connected component via a random walk from another component would be zero. Assumption 3 holds generally for any large network---for unweighted networks $p_{ij}=1/m \ll 1$ when there are many edges, while for weighted networks it holds so long as the weight of one or a few edges doesn't constitute a large fraction of the total weight of all edges.

As accumulation of memories is stochastic, a random walk of length $t$ may result in a distribution of possible memory matrices $P(\mathbf{M},t)$. However, if the assumptions 1 and 2 hold, accumulation of memories of each edge is independent and follows a distribution $P_{ij}(k,t)$, where we used $k\in\{0,1,2,\dots\}$ as an index of the distribution to simplify notation. Since the only event that can happen to an edge is addition of a count, the evolution of the distribution follows a relatively simple master equation:
\begin{align}
    P_{ij}&(k,t+1)-P_{ij}(k,t)\nonumber\\
    &=\begin{cases}
    p_{ij}(t)\left( P_{ij}(k-1,t)-P_{ij}(k,t) \right),\quad &k>0\\
    -p_{ij}(t)P_{ij}(k,t),\quad &k=0
    \end{cases};\label{eqn:master}\\
    P_{ij}&(k,0)=\delta_{k,0},
    \label{eqn:initcond}
\end{align}
where the first equation expresses the dynamics of the distribution and the second one expresses the initial condition (the edge starts with no memories).

In general, the expression~\ref{eqn:master} is an infinite system of coupled equations. However, we can attempt to solve them with the following ansatz:
\begin{align}
    P_{ij}(k,t)=\frac{\left( E_{ij}(t) \right)^k e^{-E_{ij}(t)}}{k!},
    \label{eqn:poisson}
\end{align}
which is the Poisson distribution with a single, yet-to-be-determined time-dependent parameter $E_{ij}(t)$. Substituting $E_{ij}=0$ recovers the initial condition of Eq.~\ref{eqn:initcond}. It remains for us to show that the Poisson distribution holds at all times and to find the growth law for $E_{ij}$.

We first simplify the right hand side of Eq.~\ref{eqn:master} as follows:
\begin{align}
    p_{ij}(t)&\left( P_{ij}(k-1,t)-P_{ij}(k,t) \right)\nonumber\\
    &=p_{ij}(t)P_{ij}(k,t)\left( \frac{k}{E_{ij}(t)}-1 \right),
\end{align}
where we used the functional form of the Poisson distribution. We now observe that the whole expression is proportional to $p_{ij}\ll 1$ by assumption 3. Thus the probability distribution cannot change too rapidly in a single step. We therefore approximate the finite difference on the left hand side with a derivative:
\begin{align}
    P_{ij}(k,t+1)&-P_{ij}(k,t)\approx \Delta E_{ij} \frac{\partial P_{ij}(k,t)}{\partial E_{ij}}\nonumber\\
    &=\Delta E_{ij} P_{ij}(k,t)\left( \frac{k}{E_{ij}(t)}-1 \right),
\end{align}
where we recovered an identical $k$- and $t$-dependent expression in the brackets that can be cancelled out.

From the transformed left and right hand sides of the master equation, we can now recover the simple dynamics of the Poisson distribution parameter:
\begin{align}
    \Delta E_{ij}=&p_{ij}(t);\\
    E_{ij}(t)=&\sum\limits_{t'=1}^{t}p_{ij}(t'),
    \label{eqn:exposuresum}
\end{align}
where we can now call the quantity $E_{ij}(t)$ \emph{integral exposure} of the edge $(i,j)$. We showed that so long as the three assumptions hold, the distribution of memories of the edge follows the Poisson shape (Eq.~\ref{eqn:poisson}); the process of memory accumulation is a Poisson process. For the non-existent edges of the network $p_{ij}(t)=0$ for all $t$, and thus they never accumulate any memories.

For a dynamic network, the edge visitation probability is a function of 
the evolution time $\tau$, which is related to the random walk time via dilation $t=D\tau$. For long times and smooth evolution we can approximate the sum Eq.~\ref{eqn:exposuresum} with an integral that makes the change of variables much more straightforward:
\begin{align}
	E_{ij}(t)= \int\limits_0^t p_{ij}(\tau')dt'=D\int\limits_0^{t/D} 
	p_{ij}(\tau')d\tau'=D\mathcal{E}_{ij}(t/D),
	\label{eqn:exposure_temp}
\end{align}
which recovers the formula from the main text. For time-dependent networks, $p_{ij}(\tau)=0$ for some edges for some part of the network evolution. Thus new edges cannot be accumulated but exposure does not decrease. Exposure thus accounts for both edge weight and time dependence.

\section{Poisson process calculus and coarse-graining}
The above theory was developed assuming that the elementary events of interest 
are visitations of edges. However, for other applications we might be 
interested in the memory of a \emph{group} of edges $(i,j)\in g$. Since the 
visitations are independent, we can just compute the distribution of memory 
counts for the whole group. Poisson processes are additive, regardless of the 
parameter; that is:
\begin{align}
	\sum\limits_{ij\in g}\textsf{Pois}(E_{ij}(t))= 
	\textsf{Pois}\left(\sum\limits_{ij\in g}E_{ij}(t)\right),
	\label{eqn:coarsegrain}
\end{align}
where the equality states that the left and right sides of the equation have identical distributions. Drawing pseudorandom numbers from distributions on a computer is typically a computationally expensive operation (see Section~\ref{sec:benchmark} below), while addition is cheap. Drawing a realization from the left hand side of Eq.~\ref{eqn:coarsegrain} requires doing the expensive operation once for each edge, while drawing a realization from the right hand side requires the expensive operation only once at all. The existence of the Eq.~\ref{eqn:coarsegrain} thus promises a significant quantitative and computational benefit.

The benefits of the \emph{coarse-graining} Eq.~\ref{eqn:coarsegrain} are not only quantitative, but qualitative and conceptual as well. Since the group $g$ can be defined arbitrarily, we can use the expression to compute the exposure $E_g$ of different groups and attach group-specific meaning to it. The choice of group is equivalent to the choice of an order parameter in conventional Boltzmann statistical mechanics \cite{goldenfeld}. We show two particular examples of the group $g$ below, but emphasize that other options are possible.

One particular choice of the group is all edges $(i,j)$ that connect to a particular node $j$. Since traversing any of those edges is identical to visiting the node $j$, we can use the coarse-graining formula~\ref{eqn:coarsegrain} to find the distribution of memories of visiting a node, which provides a view of the learned network complementary to the edges. We can thus use the \emph{edge} exposure $E_{ij}(t)$ 
to compute the \emph{node} exposure:
\begin{align}
	K_j(t)&\equiv \sum\limits_i E_{ij}(t)=\sum\limits_{t'=0}^t \pi_j(t');\\
	M_j&\sim \textsf{Pois}(K_j(t)),
\end{align}
where $\pi_j$ is the instantaneous steady-state visitation probability of a node.

Because the node exposures are sums of non-negative edge exposures, they are 
typically larger than edge exposures. From the exposures, we can compute the probability of visiting a node with Eq.~\ref{eqn:Pvisit}, which would grow with time much faster than the probability of visiting edges. For a network that is either unweighted or has a narrow edge weight distribution, edge exposures are $E_{ij}=\order{1/m}$. For a network with a narrow degree distribution, node exposures are $K_j=\order{1/n}$. A random walk in such networks would visit all the nodes in $\order{n \ln{n}}$ time and all the edges in $\order{m \ln{m}}$ time, consistent with prior results \cite{asztalos2010network,maier2017cover}. For heterogeneously structured networks, which are our focus here, the exploration is non-uniform and full exploration of the entire network can take much longer than it does in homogeneously structured networks.

Another choice of the group is to just include all edges $(i,j)$ and thus find the total exposure and the number of memories of the random walker. By analogy with Boltzmann statistical mechanics we can call such a sum the partition function:
\begin{align}
    \mathcal{Z}(t)&=\sum\limits_{ij}E_{ij}(t)=\sum\limits_{t'=0}^t \sum\limits_{ij}p_{ij}(t')=t;\\
    M&\sim\textsf{Pois}(t).
\end{align}
Thus regardless of the dynamics of network evolution the total number of memories fluctuates around the walk length $t$, as would be expected for a canonical ensemble.

Along with the distributions of memories of a group of edges, we can compute averages or observables over the distribution. A common quantity of interest is a binary variable of edge visitation, i.e. whether there are any memories of that edge. We compute the probability of visitation from the Poisson distribution (Eq.~\ref{eqn:poisson}):
\begin{align}
    P(M_{ij}>0,t)=1-P(M_{ij}=0,t)=1-e^{-E_{ij}(t)},
    \label{eqn:Pvisit}
\end{align}
which we use to compute the expected visited edge fraction and the cumulative distribution functions in the two following sections, respectively.

\section{Visited edge fraction and the Jensen bound proof}
Exposure theory predicts the probability of visiting any particular edge in Eq.~\ref{eqn:Pvisit}. By averaging this probability over all edges, we compute the expected fraction of all edges visited by time $t$:
\begin{align}
	V(t)=1-\frac{1}{m}\sum\limits_{ij\in A_{ij}}e^{-E_{ij}(t)},
	\label{eqn:Vfraction}
\end{align}
where the sum runs over the edges of nonzero weight. Here $\phi(x)=e^{-x}$ is a convex function, and thus we can apply Jensen's inequality:
\begin{align}
	1-\frac{1}{m}\sum\limits_{ij\in A_{ij}}e^{-E_{ij}(t)}\leq 
	1-e^{-\frac{1}{m}\sum\limits_{ij\in A_{ij}}E_{ij}(t)}.
\end{align}

In order to simplify the expression, on the right hand side we exchange the summation order over $(i,j)$ and $t$. Since at every time step the network as a whole gets exactly 1 unit of exposure, over $t$ time steps it gets $\mathcal{Z}(t)=t$ units of exposure, turning the bound into:
\begin{align}
	V(t)\leq 1-e^{-t/m}.
	\label{eqn:Vbound}
\end{align}

Recall that for an unweighted network, every edge has the same exposure $E_{ij}=t/m$. For such a network, a direct computation of the sum Eq.~\ref{eqn:Vfraction} results in the right hand side of Eq.~\ref{eqn:Vbound}. In other words, random walks on unweighted networks exactly and uniquely saturate the Jensen bound of exploration.

\section{CDF computation}
Here we show how to compute the shape of the cumulative distribution functions 
(CDF) shown in Fig.~3b,c of the main text. Since the computation proceeds 
identically for both edge weight $A_{ij}$ and filtration $F_{ij}$, we just 
denote the edge variable with $X_{ij}$. We define the \emph{non-decreasing 
ordering} of edges $(i,j)\to q$ such that $X_{q_1}\leq X_{q_2}, \,\forall(q_1<q_2)$. Whenever multiple edges have the same value of $X$, their relative order is arbitrary. Every edge has a probability $P_k$ of being visited in a particular process. The CDF can be drawn as a parametric curve in index $q\in[0,m]$ with the coordinates along the axes equal to:
\begin{align}
	\left( X_q,\; \sum\limits_1^{q} P_{q'}/\sum\limits_1^{m} P_{q'} \right).
	\label{eqn:CDF}
\end{align}

In order to draw this curve, we need to find the probability $P_k$ for each of 
three cases: the original textbook network, the stochastic simulation, and the 
exposure theory prediction. They are computed as follows:
\begin{align}
	P_q=\begin{cases}
		1,&\text{textbook}\\
		[M_q^r>0],&\text{stochastic}\\
		(1-e^{-E_q(t)}),&\text{exposure}
	\end{cases},
\end{align}
where we used the \emph{same ordering} $(i,j)\to q$ determined from the values of $X_{ij}$. For the original network, every edge is present, so on the vertical axis the points of the curve Eq.~\ref{eqn:CDF} are equally spaced. For the stochastic simulation, we use the elements of the memory matrix $M_{ij}^r$ from the run replica $r$. For the exposure prediction, the probability of each edge visit is 
given by Eqn.~\ref{eqn:Pvisit}. At very long runtimes, every edge would be 
visited at least once, so both the stochastic and the exposure expressions would approach 1.

\section{Exposure predicts memory counts}

Another way to directly test the predictions of exposure theory is to compare the statistics of the accrued memories of each edge. The number of memories $M_{ij}$ is always a non-negative integer. From stochastic simulations, we compute the mean and standard deviation of memories of every edge $\expval{M_{ij}},\sigma_{M_{ij}}$. 
From exposure theory, the number of memories has a Poisson distribution, which 
has the mean and variance (first and second cumulants) equal to the parameter:
\begin{align}
	M_{ij} \sim&\textsf{Pois}(E_{ij}(t)).\\
	\expval{M_{ij}}_c=\expval{M_{ij}^2}_c&=\sigma_{M_{ij}}^2=E_{ij}(t)
\end{align}

\noindent The exposure of edges varies over many orders of magnitude: some edges surely 
get many memory counts, while others barely get any. The threshold for edge 
discovery, as discussed in the main text, is $E_{ij}(t)\simeq 1$. If the 
``typical'' memory counts fit in the range of Poisson mean $\pm$ standard 
deviation, for $E_{ij}<1$ this range starts including the value of 0: 
fluctuations in counts become larger than the mean.

\begin{figure}
	\includegraphics[width=\columnwidth]{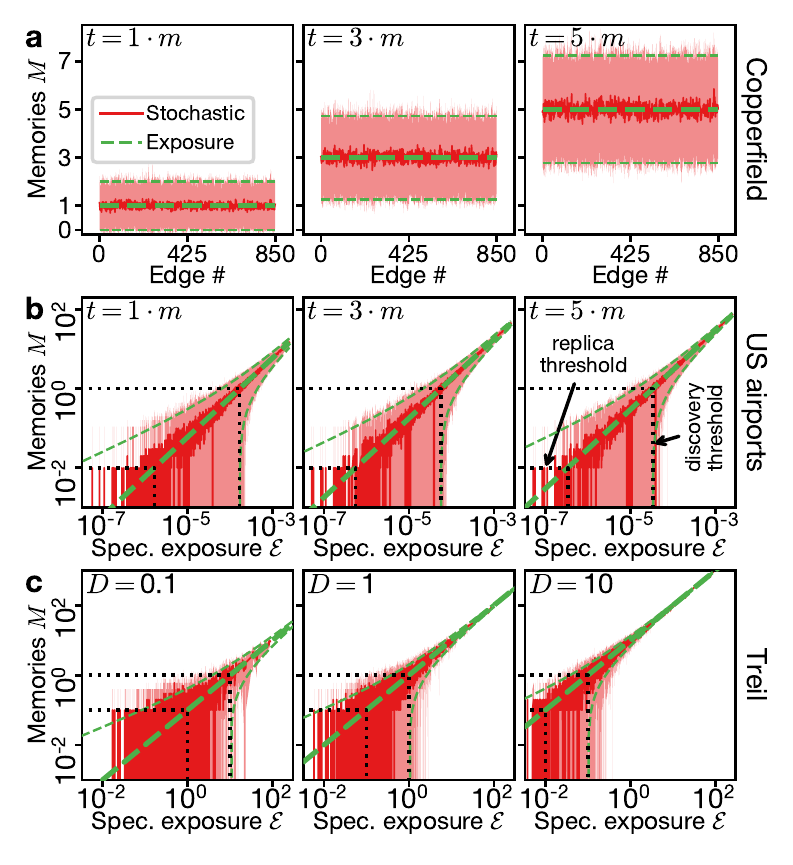}
	\caption{\textbf{Memories accrued for every edge of the networks at different time 
	$t$ or dilation $D$.} Networks include (a) David Copperfield, (b) US airports, and (c) Treil linear algebra textbook. For (a), edges are ordered arbitrarily, whereas for (b-c) edges are ordered by specific exposure $\mathcal{E}$. For (b,c), the black dashed contours indicate the discovery threshold $M=1$ and the replica threshold $M=1/R$. The red solid line and shading are stochastic mean$\pm$std, respectively; the thick and thin green dashed lines are the exposure mean and the $\pm$std range, respectively.}
	\label{fig:memories}
\end{figure}

We compare the memory count range between the stochastic simulations and exposure theory (Fig.~\ref{fig:memories}). For the unweighted Copperfield network (panel a), all edges have identical exposure and thus we expect all edges to have identical memory counts. We compute the memory counts at times $t=\{m,3m,5m\}$, corresponding to 1, 3, and 5 visits per edge on average. The resulting memory count distribution across $R=100$ replicas is indeed flat across the edge number, with a wide and uniform standard deviation range.

For the weighted US airports network, the exposure of each edge is proportional to time $E_{ij}=t\mathcal{E}_{ij}$, but also varies over almost 5 orders of magnitude (Fig.~\ref{fig:memories}b). Across the whole range of 
exposure, the stochastic memory count range follows the shape predicted by 
the Poisson distribution. The standard deviation range dips down to zero 
right at the discovery threshold $\mathcal{E}_{disc}=1/t$, but for longer times $t$ this discovery threshold moves to the left, so that edges with lower relative weight are discovered. For the US airport network we also have a second threshold associated with the finite number of $R=100$ stochastic replicas. Edges with exposure of $E=1/R=10^{-2}$ are expected to be seen only once in 100 replica runs; that is, they are rare events. Standard stochastic sampling that we use here has limited capacity to estimate the frequency of such rare events, but the exposure theory prediction is valid for arbitrary values of $E$. As the simulation runtime changes from $t=1\cdot m$ to $t=5\cdot m$, the replica threshold $\mathcal{E}_{repl}=1/Rt$ moves to the left as well.

For the  weighted and temporal Treil network, the edge exposure accumulated by the end of the textbook is proportional to the dilation $E_{ij}=D\mathcal{E}_{ij}$ (Fig.~\ref{fig:memories}c). Due to longer runtimes of simulation, we only simulated $R=10$ replicas here. Similarly to the US airports network, the stochastic memory count range follows the shape predicted by the Poisson distribution, with more noise due to fewer replicas. Since the difference in subsequent dilation $D$ values is a factor of 10, the discovery threshold $\mathcal{E}_{disc}=1/D$ shifts much more significantly between the panels than it did for the US airport network.

In conclusion, the Poisson process of memory accrual is accurate for 
describing not only the binary edge visitation probability, but also the distribution of the number of visits. The relative fluctuations in the number of memory counts $M$ fall off as $1/\sqrt{M}$ for high exposure, as expected for the Poisson distribution.

\section{Random walk correlation time}
\label{sec:correlation}
\begin{figure}
	\includegraphics[width=\columnwidth]{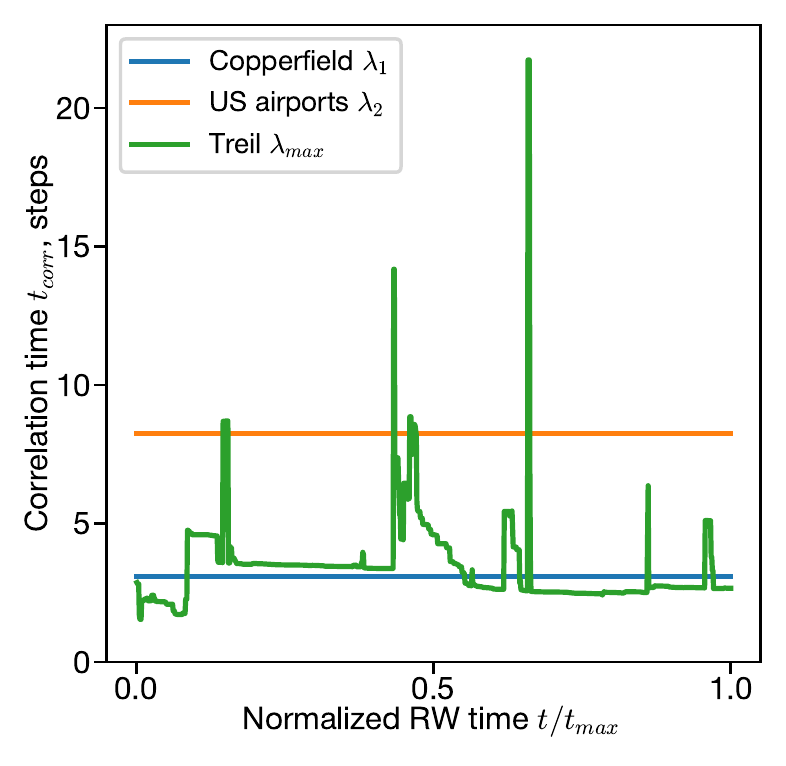}
	\caption{\textbf{Correlation times for the three networks across the normalized random walk 
	time.} Networks include David Copperfield (blue), US airports (orange), and Treil linear algebra textbook (green) 
	networks.}
	\label{fig:correlation}
\end{figure}

As a random walker explores the network, its probability of ending up on a particular node $i$ in exactly $t$ steps can be expressed as \cite{masuda2017random}:
\begin{align}
	p_i(t)=\sum\limits_{k=0}^{n-1} a_k v_i^{k} \lambda_k^t,
\end{align}
where $\lambda_k$ are the eigenvalues of the transition matrix $P(j|i)$, 
$v_i^k$ is the $i$th component of the $k$th left eigenvector, and $a_k$ are
coefficients that depend on the initial conditions. For connected graphs, over long times the probability distribution approaches the asymptotic distribution given by the top left eigenvector $v_i^0=\pi_i$. How long does that decay take?

By the Perron-Frobenius theorem, the transition matrix has a largest eigenvalue of 
$\lambda_0=1$, and all other eigenvalues are smaller or equal by absolute value $\abs{\lambda_k}\leq 1$. The presence of multiple eigenvalues of $\lambda_k=1$ 
indicates the existence of multiple network components. All eigenvalues smaller than 1 by absolute value set up the hierarchy of timescales equal to $t_{k}\simeq -1/\ln(\abs{\lambda_k})$ and commonly referred to as mixing, relaxation, decay, or correlation times. The corresponding eigenvector indicates which nodes are involved in the relaxation mode---whether just a few or many. Typically the second largest eigenvalue $\lambda_1$ is taken to compute the mixing, i.e. correlation time. If the correlation time is close to the one step of random walk $t_{corr}\simeq 1$, the subsequent nodes visited are effectively independently sampled from the instantaneous steady state distribution $\pi_i$, and thus assumptions 2 and 3 of exposure theory hold. However, since the three networks we study have a different nature, we re-examine the correlation time for each of them separately
(Fig.~\ref{fig:correlation}).

The Copperfield network is connected and unweighted. It has exactly one 
eigenvalue of 1, with the next eigenvalue by absolute value equal to $\lambda_1\simeq -0.7235$. Note that the eigenvalue is negative, which corresponds to the underlying disassortative (nearly bipartite) structure of the graph. The network originates from the adjacency of nouns and adjectives in the text of an English novel, where each part of speech is more often paired with the opposite part; that is, there are more noun-adjective adjacencies than noun-noun and adjective-adjective adjacencies \cite{newman2006finding}. The negative eigenvalue implies that correlations decay in an oscillatory manner, but we can compute the mixing time nonetheless to be $t_{corr}\simeq 3.09$. Since the correlation time is close to 1 and much smaller than the typical random walk times we consider $t_{corr}\ll t\sim m=850$, the assumptions of exposure theory hold for the Copperfield network.

The US airports network is also connected, but has a wide range of weights. 
After the first eigenvalue of 1, it has the second eigenvalue of 
$\lambda_1\simeq 0.9976$, corresponding to $t_{corr}\simeq 419$ steps, a very 
large number. However, the corresponding second left eigenvector is entirely 
localized to 5 nodes with the lowest weighted degree (also known as the nodal \emph{strength}). It takes over 400 steps for the random walk to discover those nodes, but they are not informative of how the rest of the network is explored. In order to estimate the speed of exploration of the rest of the network, we turn to the third eigenvalue of $\lambda_2\simeq 0.8857$, corresponding to the timescale $t_{corr}\simeq 8.24$. The third eigenvector is delocalized across the network, corresponding to broader mixing. This correlation time is larger than for the Copperfield network, but still much smaller than the typical random walk times $t_{corr}\ll t\sim m=5960$, so the assumptions of exposure theory also hold for the US airport network.

The Treil network, unlike the previous two, is temporal. As the network 
evolves, it changes structure and the corresponding correlation times. We thus 
compute the instantaneous spectrum of the transition matrix generated by 
normalizing the temporal adjacency matrix $\mathbf{A}(\tau)$. We rescale the 
evolution time to fit within $[0,1]$ to illustrate the dynamics. While the 
whole Treil network is connected, during the evolution it consists of a large connected component of most nodes and small disconnected components of a few nodes that only exist for several steps. In the transition matrix spectrum, 
the small components manifest through additional eigenvalues of $1$ or $-1$. Such disconnected components are quickly connected back to the main part of the network. Therefore in order to estimate the correlation time we use the 
instantaneous eigenvalue $\lambda_{max}$ which is the largest by magnitude but smaller than 1. The resulting curve in Fig.~\ref{fig:correlation} mostly hovers around 3 or 4 steps, with occasional large spikes when weak connections to new 
nodes are introduced. The lifetime of such spikes is typically shorter than 
their magnitude: before the random walker finds a new node via a weak connection, the connection becomes stronger. Outside of the short-lived spikes, the correlation time stays much smaller than the typical random walk time $t_{corr}\ll D\tau_{max}$, where $\tau_{max}=6681$ sentences for the Treil textbook. The assumptions of exposure theory hold for the Treil network as well.

The number of memories of a particular edge can be computed as a sum of increments at each time step:
\begin{align}
    M_{ij}(t)=\sum\limits_{t'=1}^t \Delta M_{ij}(t'),
    \label{eqn:memoryaccum}
\end{align}
where $\Delta M_{ij}(t')$ is a random number which is equal to 1 with low probability (when the edge $(i,j)$ is visited) and 0 otherwise. The sequence of increments is auto-correlated on the timescale $t_{corr}$. We found the correlation time on all three networks to be small $t_{corr}<10^1$, while the random walk simulations run for $10^3$--$10^4$ steps. The full sum in Eq.~\ref{eqn:memoryaccum} thus consists of many uncorrelated blocks. Moreover, in these simulations we mainly focus on the estimator of the \emph{mean} of accumulated memories, which is unbiased even for auto-correlated increments \cite{newman1999montecarlo}. In conclusion, correlation time analysis shows that exposure theory assumptions are satisfied for the three studied networks at the studied random walk lengths.

\section{Breaking exposure theory}
\label{sec:breaking}
\begin{figure*}
	\includegraphics[width=\textwidth]{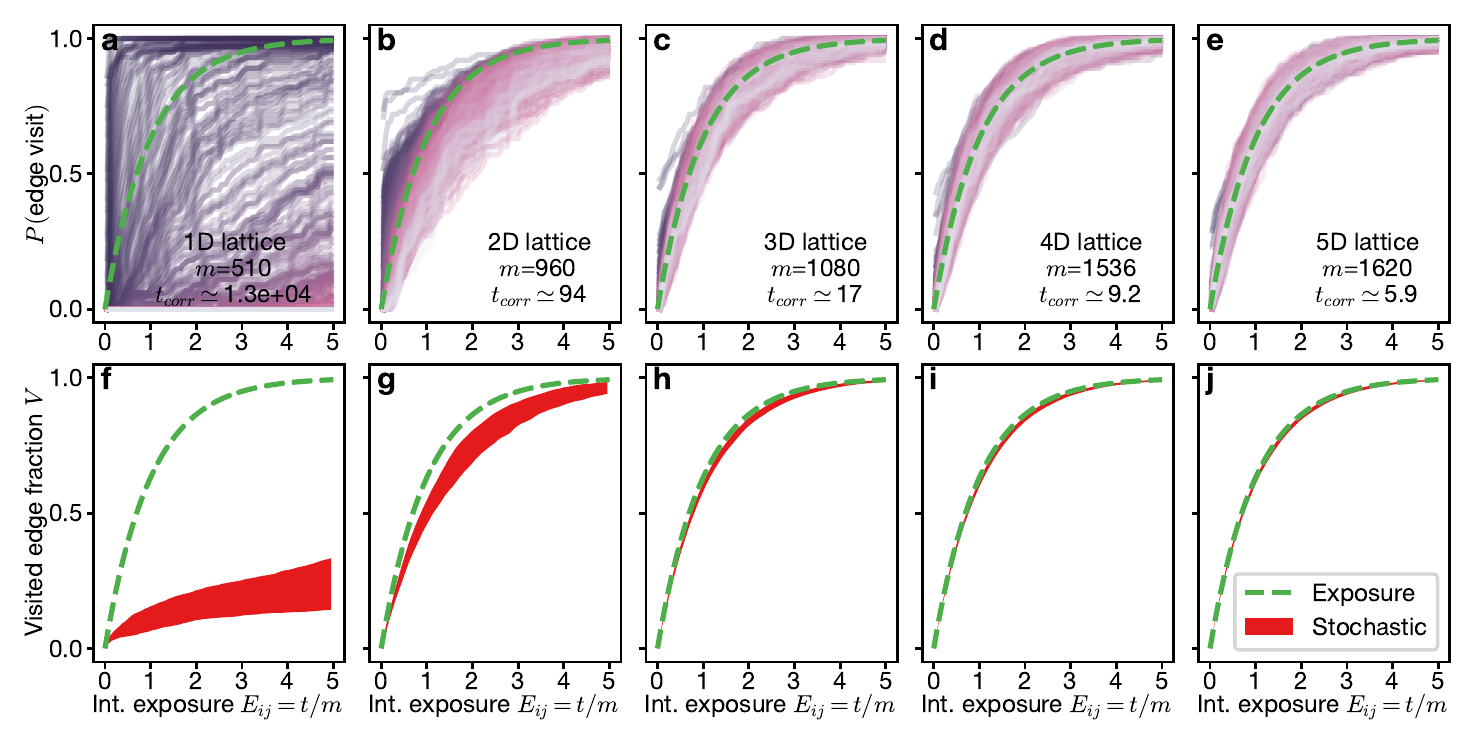}
	\caption{\textbf{Exposure theory is progressively more accurate with increasing spatial dimension.} (a-e) Probability of discovery of each network edge in lattices of increasing dimension. Line color varies from dark for closest edges to light for farthest edges. (f-j) Fraction of visited edges over time for the corresponding networks. Red shaded region corresponds to stochastic mean$\pm$std.}
	\label{fig:lattice}
\end{figure*}

\begin{figure*}
	\includegraphics[width=\textwidth]{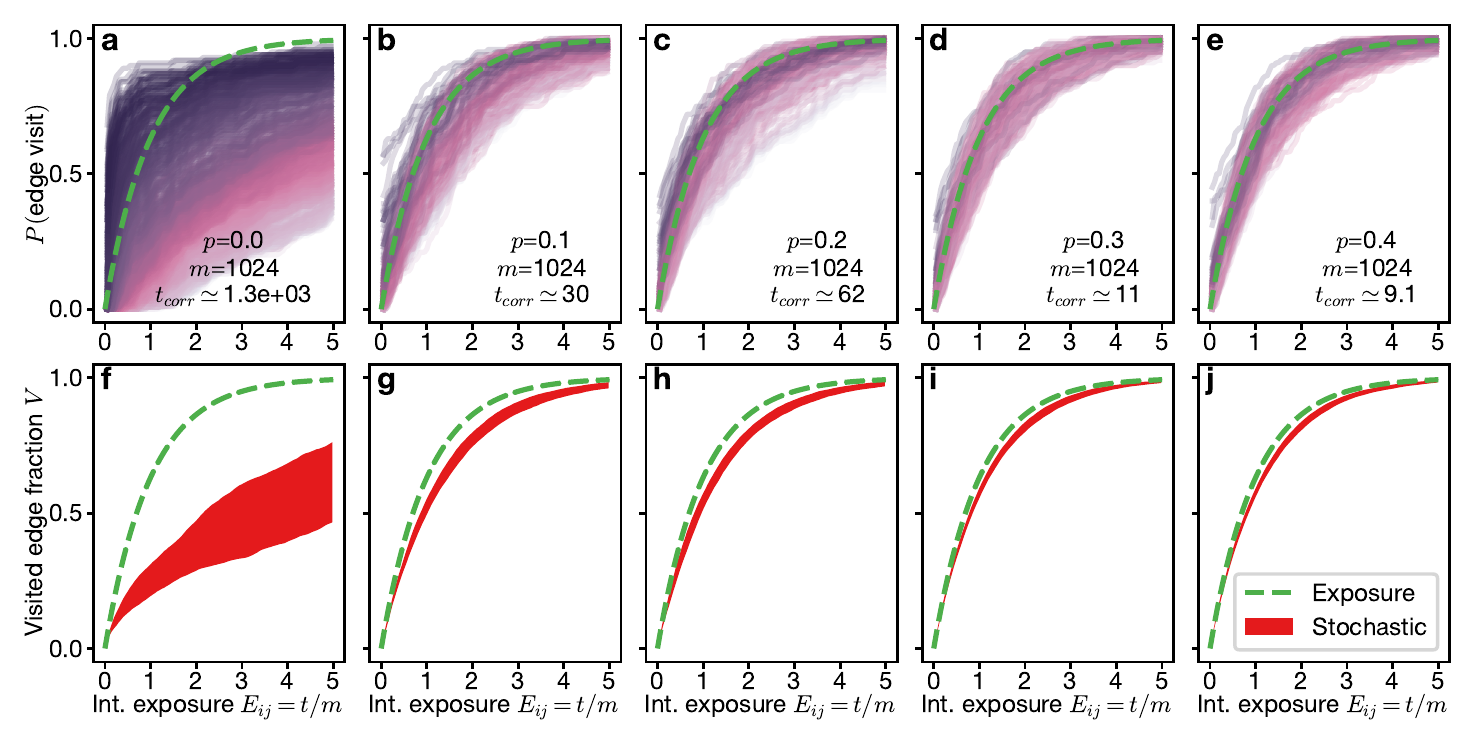}
	\caption{\textbf{Exposure theory is progressively more accurate with increasing network randomness.} (a-e) Probability of discovery of each network edge in five Watts-Strogatz networks with increasing rewiring probability $p$. Line color varies from dark for closest edges to light for farthest edges. (f-j) Fraction of visited edges over time for the corresponding networks. Red shaded region corresponds to stochastic mean$\pm$std.}
	\label{fig:WS}
\end{figure*}

In order to clarify the domain of applicability for the exposure theory, we compare its predictions to random walks on networks that explicitly break the exposure theory assumption of small correlation time, similar to Ref.~\cite{maier2017cover}. We test two groups of networks with roughly constant number of nodes. In the first group, we construct a series of five regular cubic lattices with increasing dimension and closed boundary conditions: a $n=256^1$ node 1D lattice, a $n=16^2=256$ node 2D lattice, a $n=6^3=216$ node 3D lattice, a $n=4^4=256$ node 4D lattice, and a $n=3^5=243$ node 5D lattice. In the second group we construct a series of five Watts-Strogatz small-world networks of $n=256$ nodes with $k=4$ nearest neighbor connections and increasing probability of rewiring $p\in\{0.0,0.1,0.2,0.3,0.4\}$; at each probability we consider one stochastic rewiring realization in which the network remains connected \cite{watts1998collective}. On each network, we compute 100 replicas of stochastic random walk simulations, always starting from the same node. Across both groups all networks are unweighted and undirected. Thus exposure theory has identical local and aggregate predictions in Eqs.~\ref{eqn:Pvisit}, \ref{eqn:Vfraction}. But do simulations follow the prediction?

While according to exposure theory all edges should have an identical learning curve, in stochastic simulations the curves differ significantly. For 1D and 2D lattices (Fig.~\ref{fig:lattice}a-b), the closest edges have notably higher discovery probability than exposure theory predicts, and the farthest edges have lower probability. For the 1D network, several edges have never been discovered over 100 replicas running for time $t=5\cdot m$. In contrast, for 3D, 4D, and 5D lattices, the stochastic curves closely follow the exposure prediction (Fig.~\ref{fig:lattice}c-e). We observe a similar pattern for the aggregate metric of visited edge fraction. For the 1D lattice (Fig.~\ref{fig:lattice}f) exposure theory drastically overestimates the rate of network exploration. As the lattice dimension gets higher, the standard deviation of the stochastic exploration curve decreases and the mean gets closer to the exposure prediction. As expected, exposure theory fails for low-dimensional lattices but works well for high-dimensional ones.

The pattern of prediction success is similar for the Watts-Strogatz networks. Before any rewiring occurs, local edge visitation curves strongly deviate from the exposure prediction and aggregate exploration is much slower than predicted (Fig.~\ref{fig:WS}a,f). As the network is gradually rewired, the visitation curves get more consistent and the visited edge fraction more closely follows the prediction, with smaller variance (Fig.~\ref{fig:WS}b-e,g-j). As expected, exposure theory fails for un-rewired Watts-Strogatz networks (effectively one-dimensional), but works well once the networks are randomized.

We thus showed a scenario in which exposure theory gives incorrect predictions. This scenario can be broken in two very different ways: either by constructing regular high-dimensional lattices, or by abandoning dimensional structure in favor of irregular rewiring. Both ways lead to reduction of the correlation time $t_{corr}$ (Figs.~\ref{fig:lattice},\ref{fig:WS}a-e). The drop in correlation time decreases the conditional dependence of edge visitations on the starting node, thus ensuring that a key assumption of exposure theory holds. Thus exploration of either high-dimensional lattices or random networks is qualitatively and quantitatively similar to exploration of the complex networks shown in the main text.

\section{Computational benchmark}
\label{sec:benchmark}
\begin{figure}
	\includegraphics[width=\columnwidth]{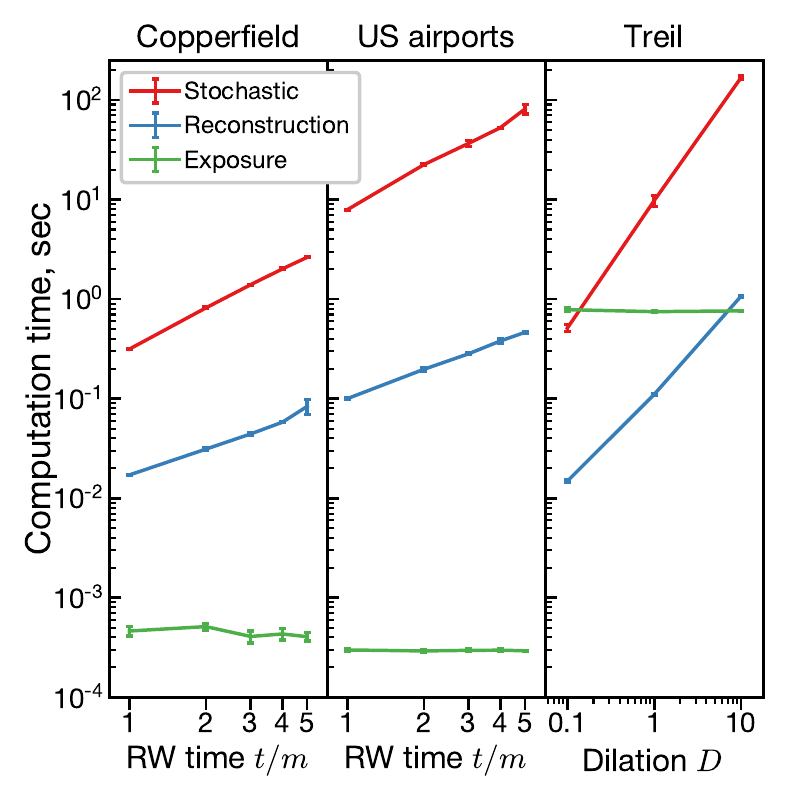}
	\caption{\textbf{Computation time of the three algorithms on each of the three networks.} Error bars indicate standard deviation over 5 runs. Note that the horizontal axis is logarithmic and spans the range $[1,5]$ for the first two networks, but a much larger range $[0.1,10]$ for the third network.}
	\label{fig:computation}
\end{figure}

In previous sections we established the accuracy of exposure theory 
predictions. However, how does exposure theory compare to stochastic simulations in terms of computational resources?

All three networks we consider in this paper are sparse; that is, the number of edges is significantly smaller than the number of possible node pairs $m\ll 
n^2$. Because of this, the adjacency matrix $\mathbf{A}$, the filtration matrix $\mathbf{F}$, the memory matrix $\mathbf{M}$, and the exposure matrix $E$ are all sparse and can be stored in $\order{m}$ memory space.

The time complexity of a stochastic random walk simulation is linear in the random walk time 
$\order{t}$. However, as we established above, the relevant timescale for 
random walks is about the number of edges $m$. Let's set $t=h\cdot m$, where 
$h$ is a small number ($1\leq h\leq 5$). The most computationally complex part of the random walk simulation is generating pseudorandom numbers to decide on the next edge to traverse, one per step. In order to save on this expense, we run the random walk simulation once for every replica and save the whole trajectory (the sequence of visited nodes). We then perform a variety of data analyses on the random walk reconstructed from the saved trajectory. While both the pseudorandom and the reconstructed trajectories give the same deterministic result (the memory matrix $\mathbf{M}$), the latter is drastically faster.

The exposure theory requires computing the exposure value for each edge. For static networks, computing the specific exposure $\mathcal{E}_{ij}$ requires dividing the weight of each edge by the sum of all weights and thus takes $\order{m}$ time. For time-dependent networks, computing the specific exposure $\mathcal{E}_{ij}(\tau)$ requires evaluating the integral in 
Eq.~\ref{eqn:exposure_temp}. For slowly-varying networks, it can be computed 
with a simple rectangle rule and thus has time complexity $\order{m\cdot 
n_{st}}$, where $n_{st}$ is the number of computational steps in time 
integration which we set to $n_{st}=10^3$. Once the specific exposure $\mathcal{E}_{ij}$ is known, it can be converted into integral exposure $E_{ij}$ through multiplying it by a scalar random walk time $t$ or dilation $D$. Finding the visitation probability via Eq.~\ref{eqn:Pvisit} or the fraction of node visits via Eq.~\ref{eqn:Vfraction} reduces to standard numerical algebra, which is optimized in modern computing packages.

We have thus established that stochastic simulation, reconstruction, and exposure computation all scale as $\order{m}$. We then run a benchmark of a Python implementation of the algorithms on a laptop computer (Intel Core i7-1065G7 @ 1.30GHz, 16Gb RAM) to get the absolute values of time, presented in Fig.~\ref{fig:computation}.

Both stochastic simulation and reconstruction scale linearly with either random walk time $h=t/m$ or dilation $D$, but reconstruction is faster than stochastic simulation by a factor of $10^1$--$10^2$. In contrast, the exposure computation runtime does not scale with either random walk time or dilation and instead takes constant time. For static networks, using exposure theory is faster than computing \emph{one replica} of stochastic simulation by a factor 
of $10^3$--$10^5$, depending on the desired random walk time and network size. For the 
dynamic Treil network, using $n_{st}=10^3$ integration steps slows down the algorithm by the corresponding factor. However, the specific exposure computation can be performed once and stored in a file for fast lookup.

The specific runtime of exposure theory computations depends on the fine 
details of implementation, programming language choice, numerical linear algebra, memory calls and data structures, and other low-level optimization. The green curve on Fig.~\ref{fig:computation} thus represents not the ultimate bound of possible performance, but merely the speed achieved by the authors of the present study. The largest gain in performance comes from not drawing pseudorandom realizations of random walk steps, but treating the probabilities as floating point numbers with fast algebra. Working directly with probability values also obviated the need to collect many stochastic random walk samples. Since getting the statistics to validate exposure predictions in this paper required between $10^1$--$10^2$ independent samples, adopting exposure theory can yield a speedup by a factor of a million.

\bibliography{biblio,biblio_SI,bibfile_CDS}